\documentclass{article}
\usepackage{amsmath,amsbsy,amsfonts,amssymb}
\usepackage{subfigure}
\usepackage{graphicx}
\usepackage{hyperref}

\usepackage[left=1.5cm,right=1.5cm,top=2cm,bottom=2cm]{geometry}

\usepackage{xcolor}

\title{Generalized  optical theorem for Rayleigh scattering approximation}

\author{  Irving Rond\'on\footnote{corresponding author: irondon@kias.re.kr}    and  Jooyoung Lee\\
	School of Computational Sciences, \\Korea Institute for Advanced Study,\\
	85 Hoegi-ro, Seoul 0245, \\Republic of Korea}

\begin{document}

\maketitle
\begin{abstract}
	A general expression for the optical theorem for probe sources given in terms of propagation invariant beams is derived.  This  expression is obtained using the far field approximation for Rayleigh regime. In order to illustrate this results is revisited the classical and standard scattering elastic problem of a dielectric sphere for which the incident field can be any propagation invariant beam.
	
\end{abstract}

\section{Introduction}
In electromagnetic theory the Optical Theorem (OT) has a very long an  interesting history  (See R. Newton \cite{RNewton} and references therein).
This physical concept of the  OT is a useful  result  in scattering theory, relating the extinction cross section of a structure  to the scattering amplitude in the forward direction \cite{Jackson,Bohren}.
In applied sciences to understand this mechanics and how the absorption and the scattering  affect wave propagation throughout a medium can help to obtain a meaningful features  of the object such as  the  characterization and its physical properties \cite{JSoares}.
\\
Recently, some interesting reviews have been  published  \cite{Gouesbet1,Gouesbet2} concerning the  development of generalized Lorenz-Mie theories, the Extended Optical theorem (EOT), using structured beam shape methods are presented, and the description of several electromagnetic effects by experiments, some applications using  T-matrix methods for structured beam illumination.
\\ 
The optical theorem has generated important and new applications, such as calculating the complex scattering amplitude  on spherical and cylindrical objects, application in acoustic backscattering \cite{PLMarston2001}, to  understand correctly the physical effects of propagation in radiation force and torque \cite{PLMarston1984}, without input nonphysical effect or based on numerical calculations  (See \cite{PLMarston2016}  and  references).
\\
A generalized optical theorem for acoustic waves has been published by Marston and Zhang \cite{Zhang2012,Zhang2013}, recently extended for arbitrary beams \cite{Zhang2019}. For electromagnetic fields \cite{Berg2008,Berg2008a, Carney1, Carney2} quantum mechanics \cite{Gouesbet2009}, in time domain, transmission lines, propagation in acoustic and electromagnetic waves in anisotropic medium \cite{Marengo1,Marengo2,Marengo3}, seismologic waves \cite{Kees}, the  manipulation of the scattering pattern using non-Hermitian particles \cite{Yun}.
\\
In the following section, a general derivation EOT applicable to any \textquotedblleft nondiffracting beams\textquotedblright using the Huygens principle in the far field approximation for Rayleigh regime. A  generalized scattering amplitude function  is presented. These results are illustrated  using a Bessel beam.

\section{Theoretical background analysis }
Let us consider a scattering particle of arbitrary form and size, with volume $V$ a boundary surface $\partial \Omega$ and, a complex permittivity 
$ \varepsilon\left( \vec{r}\right)  =\varepsilon_0\left[  \varepsilon_r'(\vec{r})+i \varepsilon_r''(\vec{r})\right] ,	
$ where $\varepsilon_0$ is the vacuum permittivity. For simplicity the medium surrounding the scattering particle is lossless and its permittivity is $\varepsilon$ and not magnetic. Let us  consider the incident arbitrary  beams as  $\vec{E}_i(\vec{r}),\, \vec{H}_i(\vec{r})$   that strikes the scattering particle,  $\vec{E}_s(\vec{r}),\, \vec{H}_s(\vec{r})$ the scattered electromagnetic field and $\vec{E}(\vec{r}),\, \vec{H}(\vec{r})$  as the total  fields \cite{Bohren,Jackson,FabrizioI,FabrizioII}.

\begin{equation}
	\label{ec:Etotal}
	\vec{E} = \vec{E}_i + \vec{H}_s, 
\end{equation}
\begin{equation}
	\label{ec:Htotal}
	\vec{H} =  \vec{H}_i + \vec{H}_s.
\end{equation}
The time-averaged power absorbed by the scattering object is given by 
\begin{equation}
	P_{a}=  -\frac{ 1}{2}\int_{\partial \Omega} \Biggl[  \vec{n} \cdot \left[ \left( \vec{E}_i + \vec{E}_s \right)\times \left( \vec{H}_i^* + \vec{H}_s^*\right)\right]  \Biggr] dS,
\end{equation}
this expression can be written in term of the Poynting vector  as 
\begin{equation}
	\label{ec:PaTotal}
	P_{a}=  -\frac{ 1}{2}\int_{\partial \Omega}  \vec{n} \cdot \left[ \vec{S}_i  + \vec{S}_s +  \vec{S}'\right]   dS,
\end{equation}
Where the first term of  Eq. \eqref{ec:PaTotal} is the incident  Poynting vector 
\begin{equation}
	\vec{S}_i =  \frac{1}{2}\mathbf{Re} \left[ \vec{E_i}\times \vec{H}^*_i \right],
\end{equation}
followed by the scattered  vector
\begin{equation}
	\vec{S}_s =  \frac{1}{2}\mathbf{Re} \left[ \vec{E}_s\times \vec{H}_s^* \right],
\end{equation}
and the cross term interaction vector, which depend in term of the initial and scatter vectors
\begin{equation}
	\vec{S}' =  \frac{1}{2}\mathbf{Re} \left[ \vec{E}_i\times \vec{H}_s^* + \vec{E}_s \times \vec{H}_i^* \right].
\end{equation}
Using these equations and the right boundary condition which depend of the specific geometry.  It is possible  to obtain  analytical and numerical solution for general scattering problem (See \cite{FabrizioI,FabrizioII} where the authors has reported a  pedagogical  tutorials). 
\\
In this letter, we follow and adapt the approach presented in ( See. Refs \cite{Tsang2000,Ishimaru1991})   for an arbitrary structured beams, where in  general the electromagnetic fields can be expressed as  $\vec{E}(r,t)= \mathbf{Re}[ {E(r)e^{-i \omega t}}]$ and $\vec{H}(r,t)= \mathbf{Re}[ {H(r)e^{-i \omega t}}]$,
where $\omega$ is the harmonic frequency of the wave (single frecuency).\\
Let us define the following  incident fields as
\begin{equation}
	\label{ec:Ei}
	\vec{E}_i = \hat{e}_i  \varphi(x,y) e^{ik_z z} e^{-i \omega t},  
\end{equation}

\begin{equation}
	\label{ec:Hi}
	\vec{H}_i =  \frac{1}{i\omega  \mu_0} \nabla \times  \vec{E}_i ,
\end{equation}
where  $\varphi(x,y)$  physically represent a  structured beam (scalar function) which satisfies the Maxwell's equations and the transversal Helmholtz equation $ \nabla_t^2 \varphi + k_t^2 \varphi=0$, in Cartesian, circular, parabolic cylindrical, and elliptical coordinates \cite{Miller,Uri,Whitaker,NietoVesperinas1991} and $k_t$ is the transversal wave vector. For simplicity the factor $e^{-i\omega t}$ is omitted.\\
In the following is considered a polarized electric field, with a complex amplitude $E_0$.
\begin{equation}
	\label{ec:Eix}
	\vec{E}_i  = \hat{e}_x  E_0\varphi(x,y)e^{ik_z z}, \\
\end{equation}

\begin{equation} 
	\label{ec:Hix}
	\vec{H}_i \approx   \hat{e}_y \frac{E_0}{\omega \mu_0 } k_z   \varphi (x,y)   e^{ik_z z},
\end{equation}
The Poynting vector for these incident fields  lies  the plane XY \cite{Bohren}.
\\
In this approximation has been assumed that  any impinging electric field  can be written as invariant structured beam using the plane wave spectrum representation \cite{Miller,Uri,Whitaker,NietoVesperinas1991}.
\begin{equation}
	\label{eq14}
	\varphi(x,y)= 
	\int_{0}^{ 2\pi}    
	A (\phi )  e^{  i k_t ( x \cos\phi + y \sin\phi) }  d\phi,
\end{equation}
where  $A(\phi)$ is an angular  function. Note that for $A(\phi)=\delta(\phi-\phi_0)$, the plane wave case is recovered,  $k_t$  is   the transversal wave vector.
This equation  has the following interpretation. A structured or non-diffracting beam is given by a superposition of multiple plane waves having transversal wave vectors $k_t$ on a circle.  Only for particular functions $A(\phi)$ on the Whittaker’s integral can be expressed analytically. For example higher order Bessel beams are defined by $A(\phi) = e^{i m \phi}$, where $m$ is the azimuthal order of the beam. Mathieu beams are defined by $A(\phi) = C(m, q, \phi) + iS(m, q, \phi)$ are the Mathieu cosine and sine functions. Weber beam are defined by $A(a,\phi)=\frac{1}{2(\pi\vert \sin \phi \vert^{1/2} )}e^{i a \ln \vert \tan \phi/2 \vert}$ in turn is divided into even and odd case \cite{Uri}.
\\
\\
After substituting Eqs. \eqref{ec:Eix} and \eqref{ec:Hix} into Eq. \eqref{ec:PaTotal} is written as
\begin{equation}
	P_{a} + P_{s} =  -\frac{ 1}{2}\int_{\partial \Omega}    \mathbf{Re} \left[ \vec{E}_i\times \vec{H}_s^* + \vec{E}_s \times \vec{H}_i^* \right] \cdot \vec{n} dS, 
	\label{ec:PaPsInvBeams}
\end{equation}

\begin{figure}[h]
	\centering
	\includegraphics[scale=0.96]{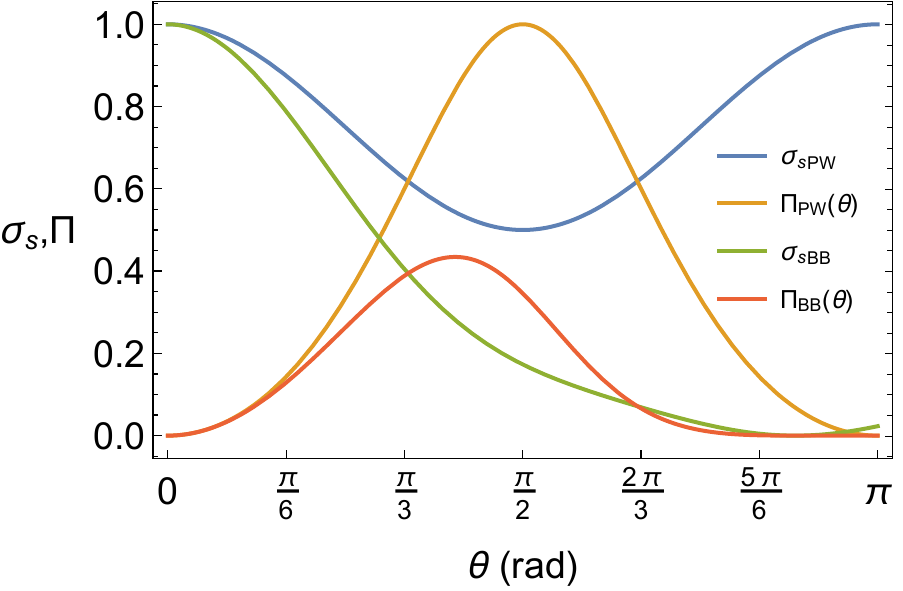}
	\caption{Differential scattering cross section  \eqref{ec:dScatBessel}  normalized and the polarization function \eqref{ec:fscatBessel}  vs scattering angle for a plane wave and Bessel beam.  In the Bessel beam case the sphere is on the beam axis and the beam order is $m = 0$}.
	\label{fig:WPlaneWBessel}
\end{figure}
\noindent
It is possible to recast  Eq. \eqref{ec:PaPsInvBeams} after some vector algebra,  as follow
\begin{equation}
	P_{a} + P_{s} = \frac{\mathbf{Re}}{2}\int_{\partial \Omega} \Biggl[  \hat{e}_x \cdot \left( \vec{n}\times \vec{H}_s \right) E_0^*
	\varphi^*(x,y) e^{-ik_z z}  +  \hat{e}_y \cdot \left( \vec{n}\times \vec{E}_s \right)  \varphi^*(x,y) \frac{ E_0^* e^{-ik_z z}}{\eta } \Biggr]dS
\end{equation}
where  $\eta = \sqrt{\mu/ \varepsilon}$ is denoted for the impedance. This is a general expression that represent from the physical point of view  the interaction between the incident and scattered fields in term of the Green function, also related with the Huygens principle\cite{Jackson,Tsang2000,Ishimaru1991}. 
Therefore, the scattered electric field in the far field can be expressed as 
\begin{equation}
	\vec{E}_s = \int_{\partial \Omega} i \omega  \mu G(\vec{r},\vec{r'})\cdot \left[\hat{n} \times \vec{H'}_s \right]
	+ \nabla \times G(\vec{r},\vec{r'})\cdot  
	\left[\hat{n}\times \vec{E'}_s \right] dS'.
\end{equation}
Alternatively, this  field in the scattered vector in the far approximation can also be expressed as 
\begin{equation}
	\label{Eq:Es}
	\vec{E}_s(\vec{r}) = \frac{e^{ikr}}{r} F(\hat{k}_i, \hat{k}_s)\cdot \hat{e}_i, 
\end{equation}
where the  function  $F(\hat{k}_i, \hat{k}_s)$ is 
called the scattering amplitude function, $\hat{e}_i$ is an unitary vector.  A similar  expression is related with the pressure in acoustic \cite{Zhang2012,Zhang2013,Zhang2019} .
\\
This function can provide itself very interesting physics in scattering phenomena. Indeed, in Refs. \cite{Marston1,Marston2} the authors proposed a method how to obtain
analytical expressions for the scattering amplitude function in order to explore an acoustic Bessel beam, later extended for Mathieu \cite{Chafiq}, and  Weber  \cite{Nebdi} waves. For instance , in order to obtain an OT expression for arbitrary beam, we use Eq. \eqref{Eq:Es} in the forward direction i.e $\hat{k}_s = \hat{k}_i$  \cite{Irving}, taking the product with  $\hat{e}_i \varphi^*(x',y') E_0^*$ , and making the integration  over  the scattered object in the far field approximation for the electromagnetic case  \cite{Mishchenko2006a,Mischenko2006b,Wolf,Carney1,Carney2, }. Also a clear derivation  for the ETO theorem for acoustic waves  using Jones lemma can be reviewed \cite{Zhang2012,Zhang2013,Zhang2019}
\begin{equation}
	\label{Eq:Green}
	\hat{e}_i  \varphi^*(x',y') \cdot \vec{E}_s  = \int_{\partial \Omega} \varphi^*(x',y') e^{-i \hat{k}_s \cdot \vec{r}'}
	\left[ \hat{e}_i\cdot (\hat{n} \times \vec{H'}_s)   + ( \hat{k}_i \times  \hat{e}_i )  \cdot (\hat{n}\times \vec{E'}_s)
	\right] dS', 
\end{equation}
Therefore, without loss generality applying  the same   physical condition as \cite{Ishimaru1991,Tsang2000,Mishchenko2006a,Mischenko2006b,Irving} using Eq. \eqref{Eq:Es} and \eqref{Eq:Green}, the optical theorem  can be written as 
\begin{equation}
	\sigma_\text{ext}=  \frac{4\pi }{k} \mathbf{Im}\left[\hat{e}_i\cdot \varphi^*(x',y')  F(\hat{k}_i, \hat{k}_i)\cdot \hat{e}_i\right]. 
\end{equation}
This expression recover the OT for the plane wave case,
\begin{equation}
	\sigma_\text{ext}  =  \frac{4\pi }{k} \mathbf{Im}\left[\hat{e}_i\cdot  F(\hat{k}_i, \hat{k}_i)\cdot \hat{e}_i\right] 
\end{equation}

\section{Generalized scattering amplitude function}
In this section, we show an application of the scattering amplitude function. Let us consider an electric field  
\begin{equation}
	\label{EcCampEsfDiel2}
	\vec{E}= \frac{3}{2 + \varepsilon_r} \vec{E}_i.
\end{equation}
this equation describe the electric field $\vec{E}$ inside a dielectric sphere  immersed in term of  the incident electric field $ \vec{E}_i $ \cite{Jackson,Bohren,Ishimaru1991}
\\
\\
In order to avoid redundancy, we put forward  the scattering amplitude function the Rayleigh regime reported  \cite{Irving}, but here written in term of Whittaker integral as

\begin{equation}
	\label{ec:Finvariant}
	F(\hat{i}, \hat{o})   =   k^2 a^3 \left( \frac{ n^2 - 1 }{n^2 +2 } \right)      
	\left[ \hat{e}_i -  \left( \hat{o}\cdot  \hat{e}_i   \right)\hat{o} \right]  e^{i k_z z }
	\int_{0}^{2\pi}  A (\phi ) e^{  i k_t ( x \cos\phi + y \sin\phi) }  d\phi,    
\end{equation}
where $n$ is the complex refraction index, $k$ is the wave vector  and $a$ is the sphere radius.
\\
The square of this equation an integration over the solid angle $d\Omega$ gives the differential scattering cross section as  
\begin{equation}
	\label{Eq:newExt}
	\frac{d \sigma_{\text{s}}}{d \Omega}= 
	\vert F(\hat{i}, \hat{o}) \vert^2 =
	k^4 a^6 \left( \frac{ n^2 - 1 }{n^2 +2 } \right)^2   I_{\text{beam}},  
\end{equation}
where
$
I_{\text{beam}}=  \left[ 1 -  \left( \hat{o}\cdot  \hat{e}_i   \right)^2 \right]   \vert   \varphi(x,y)\vert^2,   
$
is the intensity expressed in term of a scalar potential, it is related with the incident angle and observation point, if $\hat{e}_i$ is perpendicular to the scattering plane, and $\hat{o}\cdot \hat{e}_1= \sin\theta$, if $\hat{e}_i$ lies in the plane, $ \vert   \varphi(x,y)\vert^2$ is related with 
the intensity.  Using this expression several  probe fields can be used to measure the scattering cross section. 
In the following, it is denoted $\sigma_{\text{d}} \equiv  d \sigma_{\text{s}}/d \Omega$.

\begin{figure} [htbp!]
		\centering
		\includegraphics[width=7.5cm]{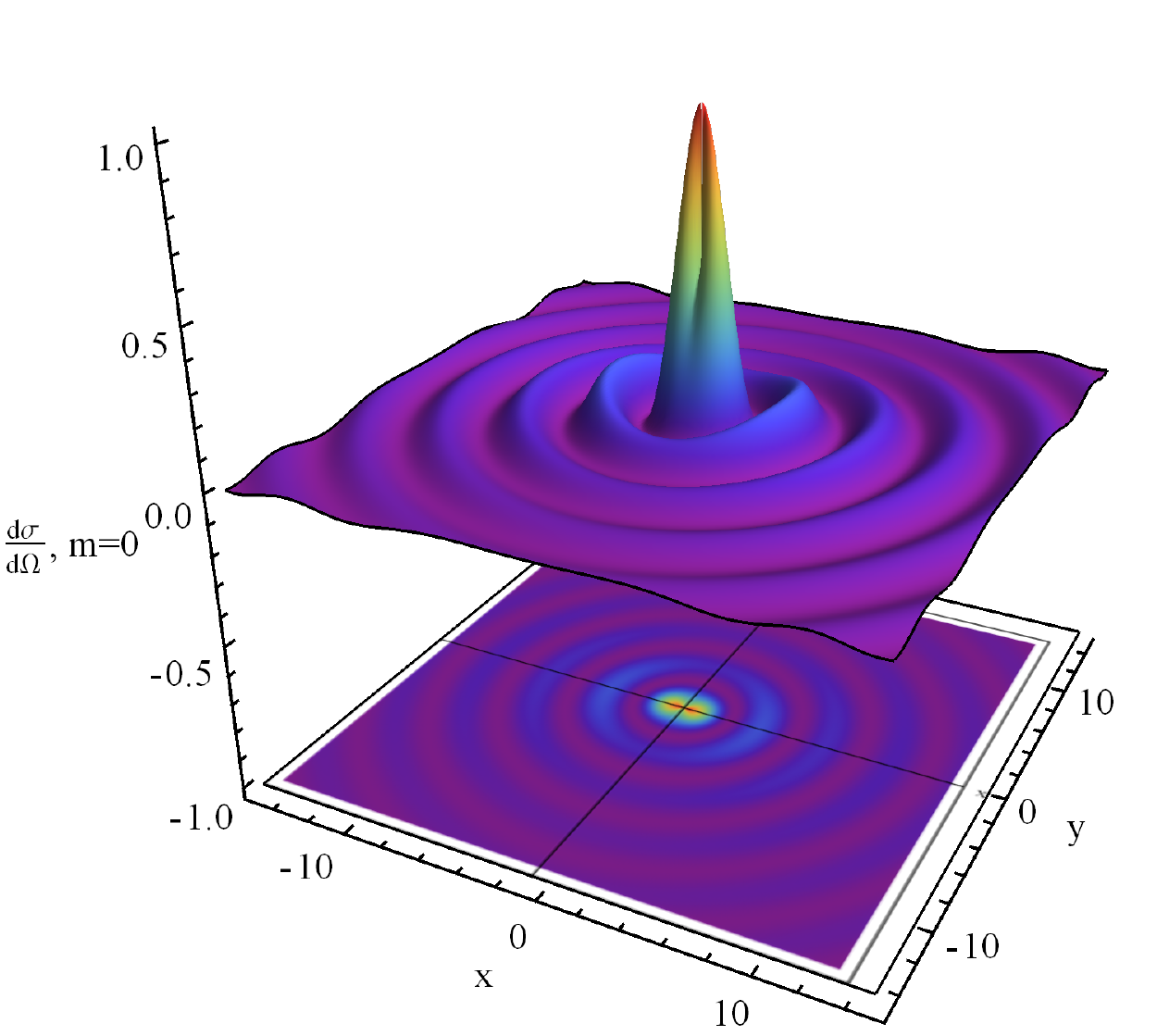}
		\qquad
		\includegraphics[width=7.5cm]{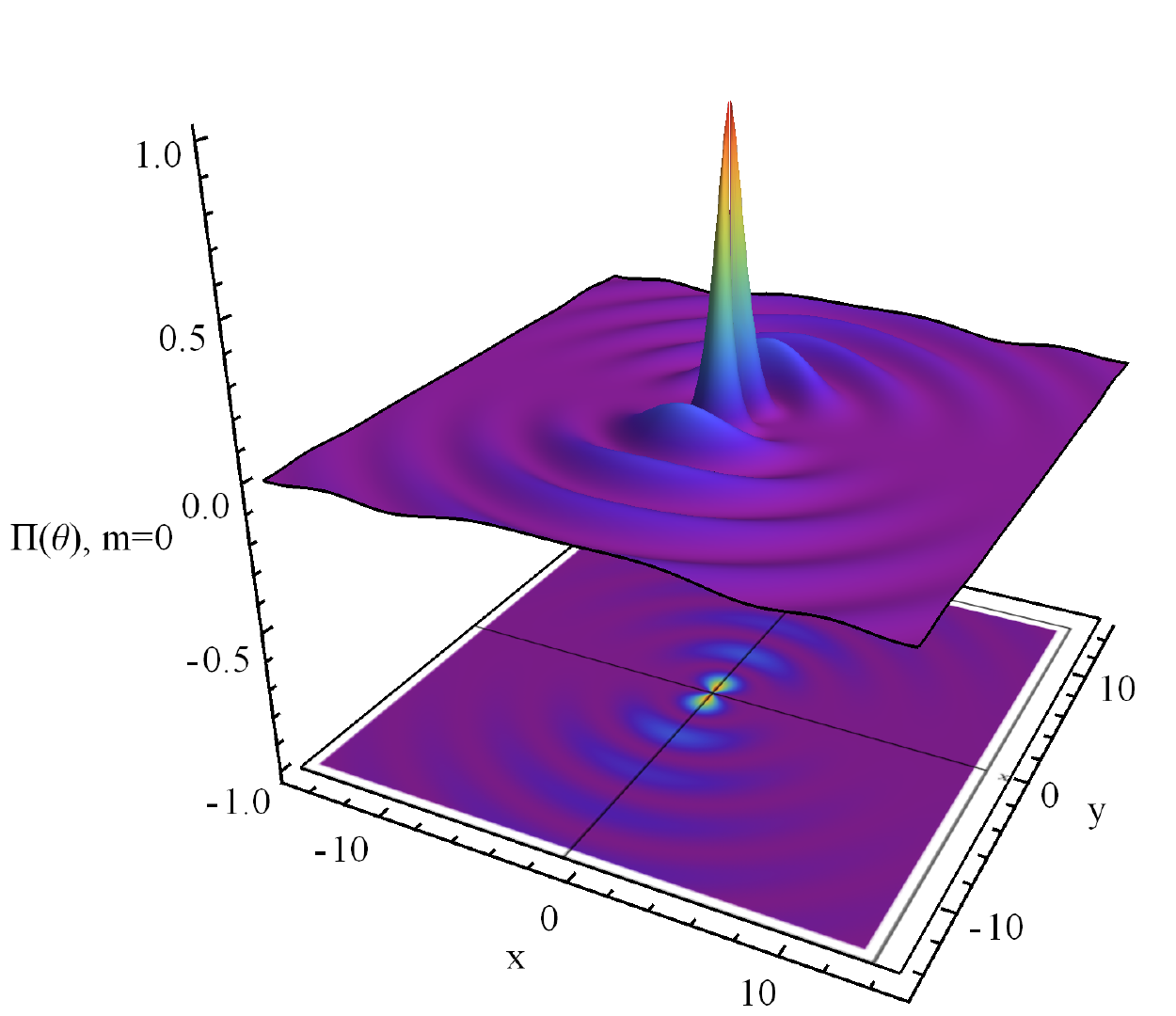}
	\caption{Normalized differential scattering cross section  and the polarization scattered function for a Bessel beam with   $m=0$,  using  the equations \eqref{ec:dScatBessel} and \eqref{ec:fscatBessel} with  $\beta= 35^{\circ}$. }
	\label{fig:3DSPBessel}
\end{figure}

\subsection{The Rayleigh scattering approximation using a Bessel beam}
In this section, we consider an incident electric field expressed using Eq.
\eqref{EcCampEsfDiel2} taking the angular spectrum as $A(\phi)=e^{im\phi}$, which  represent a linear polarized Bessel beam of $m$ order is expressed as
$\vec{E}_i(\vec{r})=  \hat{e}_i J_{m}(k_{t} r) e^{i k_z z }$. Using this result it is possible to obtain the polarization function of the scattered radiation as
\begin{equation}
	\label{ec:dScatBessel}
	\sigma_{\text{d}}(\theta)  = \frac{k^4 a^6}{2}   \left( \frac{ n^2 - 1 }{n^2 +2 } \right)^2   J_{m}^2(k_{t} r) (1 + \cos^2 \theta).
\end{equation}
In this problem,we have used the standard  scattering geometry  \cite{Shimuzu,Gordon,Drake} to relate the scattering angle  $\theta = \arccos \left( \hat{e}_i\cdot \hat{o} \right) $ to the unit vectors  $\hat{i}$  and $\hat{o}$ in a particular point \textit{P},  where the scattered radiation is observed. In the other hand,  if the incident field is unpolarized,  the differential scattering function is the average  over parallel and perpendicular incidents fields, then  $\sigma_{\text{d}}(\theta) = \frac{1}{2} \left[  \sigma^{\perp}_{\text{d}}(\theta) +\sigma^{\parallel}_{\text{d}}(\theta) \right]$,  and  $\hat{o}\cdot \hat{e}_i=0$  if $\hat{e}_i$ is perpendicular to the scattering plane and $\hat{o}\cdot \hat{e}_1= \sin\theta$ if $\hat{e}_i$ lies in the plane. Using these information we can calculate the polarization scattered function as  
\begin{equation}
	\label{ec:fscatBessel}
	\Pi(\theta) =  \frac{ \sigma^{\perp}_{\text{d}}(\theta)-\sigma^{\parallel}_{\text{d}}(\theta)  }{ \sigma^{\perp}_{\text{d}}(\theta)+\sigma^{\parallel}_{\text{d}}(\theta) }=\frac{  \sin^2\theta  }{1+ \cos^2 \theta} J_{m}^2 \left( k_t r\right).
\end{equation}

\begin{figure} [htbp!]
		\centering
		\includegraphics[width=7.5cm]{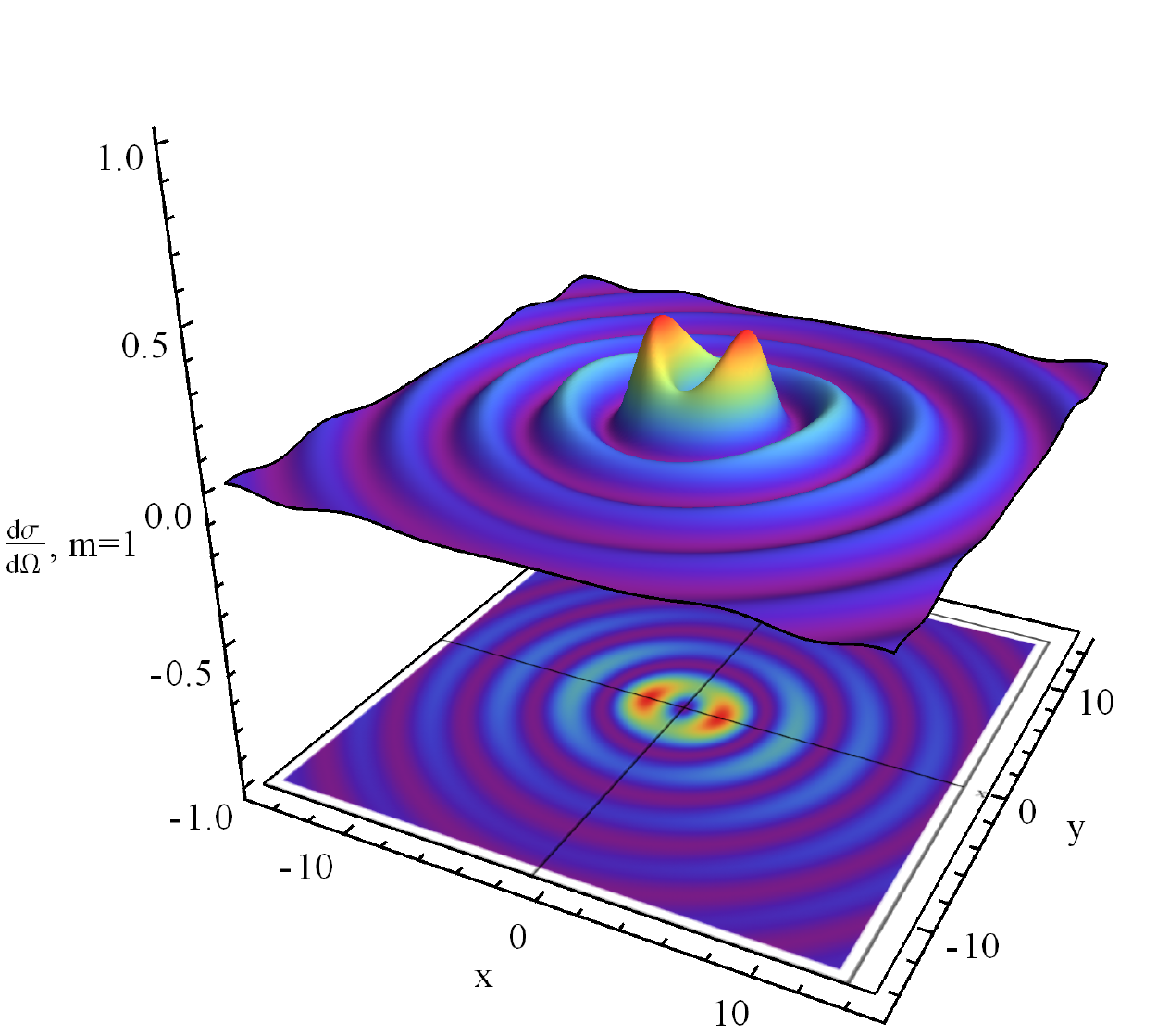}
		\qquad
		\includegraphics[width=7.5cm]{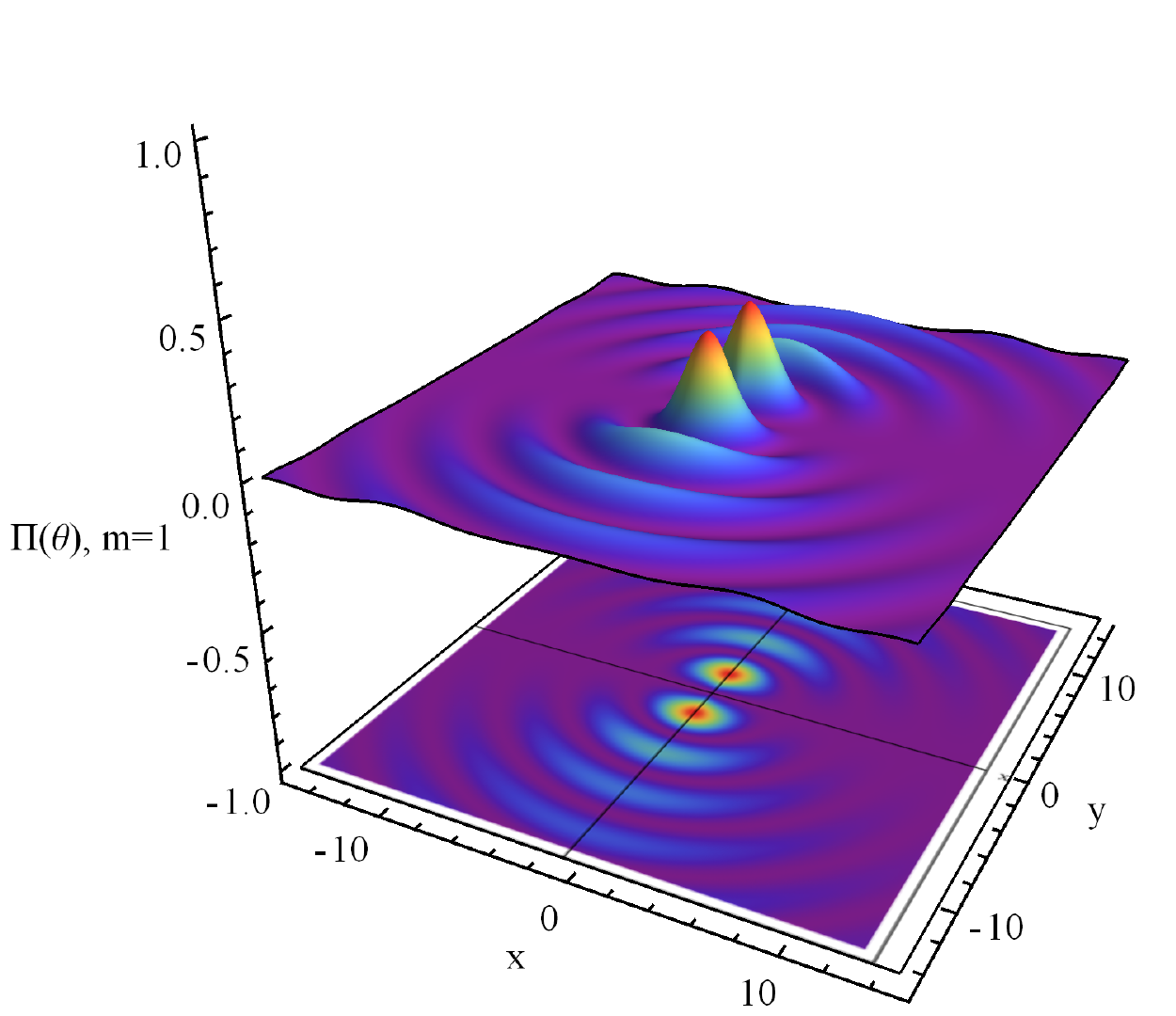}
	\caption{ Normalized differential scattering cross section and the polarization scattered function for a Bessel beam with $m=1$  and with  $\beta= 60^{\circ} $. }
	\label{fig:4DSPBessel}
\end{figure}

It is important to note, that integrating the equation Eq. \eqref{ec:dScatBessel} over $d\Omega$ is straightforward to obtain   $\sigma_{\text{s}}$ \cite{Irving}, the scattering transversal section, while the absorbed transversal section $\sigma_\text{a}$ have to be obtained with the Poynting vector \cite{Jackson,Bohren,Tsang2000,Ishimaru1991}, since in other way, this would result  in $\sigma_{\text{ext}}=0$. Note that taking $r\rightarrow 0$ into  \eqref{ec:dScatBessel} and \eqref{ec:fscatBessel} is recovered the plane wave case \cite{Jackson}. \\
\\
Finally, in order to validate our results, we show in Fig. (\ref{fig:WPlaneWBessel}) some numerical results, in which we compare the scattered radiation as function of the dispersion angle for both the differential cross section and the polarization scattered  function, considering a plane wave  and a Bessel beam with $m=0$. For simplicity, we make the analysis in  one dimension and vary the angle from  $0$ to $\pi$. For the Bessel beam case, $k_{t}= k \sin \beta$ is the transversal wave vector, and  $\beta$ is the value of the half-cone angle.\\
In addition, this model reproduce the expected behavior for a plane wave \cite{Jackson}. For a plane wave case, the  differential cross section present a minimum at $\pi/2$ and the polarization scattered function a maximum at  $\pi/2$ See  Figure 3,  these angles the scattered radiation is linear polarized,  this effect  was proposed by Rayleigh to explain the blue sky \cite{Jackson,Gordon,Shimuzu}. At these same values  for a zero Bessel beam, the differential cross section shows a reduction respect to the plane wave case, with a maximum around $\pi/3$, while the scattered polarization function shows a clear decreasing of the scattered radiation.\\
In Figure  \eqref{fig:3DSPBessel},  we show the three dimensional behavior  of the differential cross section  $\left(  \sigma_s \right) _{\text{BB}}$ and the polarization scatter function $\Pi(\theta)_{\text{BB}}$ with  $m=0$  and  $\beta = 35^{\circ}$ for a Bessel beam.  In addition, to illustrate how the scattering cross section and the polarization scattered function change with the variation of the incident field, we show in Figure  \eqref{fig:4DSPBessel} the behavior of these quantities for  $m=1$ and  $\beta=60^{\circ}$. Bi-dimensional scattered  pattern  were reported using only a zero order Bessel EM scattering in a dielectric sphere \cite{Irving} where the author compare the incident Bessel beam as a function of the  impinging angle $\beta$.

\section{Conclusions}
A general optical theorem for any arbitrary beam was derived, using the amplitude scattering function and the Huygens principle in  the far field approximation. the presented ordinary form of the  optical theorem  renders the particular case for free space  derived in previous works \cite{PLMarston2016,Zhang2012,Zhang2013,Zhang2019}. A general representation for  the extinction in Rayleigh scattering regime was studied and the effect for a  linearly polarized Bessel beam of $m$ order as a function of  the incident impinging angle $\beta$.   From the physical point of view this method  Eq. \eqref{Eq:newExt} can be extended using another spectral beam wave representation  \cite{Hugo2014}, it allows to study waves such as X waves, Airy, Frozen waves, among others.
It would be interesting to explore  as was stated \cite{Zhang2019} the physics of differential cross section of scattering $d\sigma_{\text{s}}/d\Omega  $ and  for the extinction $ d\sigma_\text{ext}/d\Omega$, the geometrical features and analogies between acoustic and electromagnetism.
Several application such as  the Rayleigh scattering of nanoparticles and optical forces \cite{LGong} using focused femtosecond laser pulses are in development. Applications  measuring  the optical extinction  has been recently experimentally  explored using a  radial polarized cylindrical beams  \cite{Krasavin}.

\end{document}